\documentclass[a4paper,10pt]{article}
\usepackage[english]{babel}
\usepackage[T1]{fontenc}
\usepackage[latin1]{inputenc}
\usepackage{indentfirst}
\usepackage[dvips]{graphicx}
\begin{document}
\title{
Quantized Transport in Two-Dimensional Spin-Ordered Structures}

\author{ Ilaria Campana$^1$, Giancarlo Jug$^2$ and Klaus Ziegler$^1$ \\
$^1$Institut f\"ur Physik, Universit\"at Augsburg, D-86135 Augsburg 
(Germany) \\
$^2$Dipartimento di Fisica e Matematica, Universit\'a dell'Insubria, 
22100 Como (Italy)}

\maketitle

\begin{abstract}
We study in detail the transport properties of a model of conducting
electrons in the presence of double-exchange between localized spins arranged
on a 2D Kagome lattice, as introduced by Ohgushi, Murakami and Nagaosa
(2000). The relationship between the canting angle of the spin texture
$\theta$ and the Berry phase field flux per triangular plaquette $\phi$ is
derived explicitly and we emphasize the similarities between this model and
Haldane's honeycomb lattice version of the quantum Hall effect (Haldane, 1988).
The quantization of the transverse (Hall) conductivity $\sigma_{xy}$ is
derived explicitly from the Kubo formula and a direct calculation of the
longitudinal conductivity $\sigma_{xx}$ shows the existence of a
metal-insulator transition as a function of the canting angle $\theta$ (or
flux density $\phi$). This transition might be linked to that observable in
the manganite compounds or in the pyrochlore ones, as the spin ordering
changes from ferromagnetic to canted.
\end{abstract}


\section{ Introduction } 

A fascinating and relatively common problem in solid-state physics is the
motion of electrons in a lattice structure where localized spins are
present, possibly in an organised form. The scattering of the electrons by
the spins represents a complex physics problem for the theoretical
investigator, especially when (like it seems to happen for the high-$T_c$
cuprate superconductors) the spin and charge degrees of freedom are attached
to the same particles -- a situation which we do not consider. In a
simplified picture we assume first of all that the spins are localised and
interacting, but that the mobile electrons do not interact with each other
(or interact weakly and give rise to independent quasiparticles) and are
separate entities from the electrons that produce the magnetic ions. In this
case the individual electrons (or quasiparticles) experience the localized
spins as an effective local magnetic flux (M\"uller-Hartmann and Dagotto, 1996). This flux produces
a Berry (or Peierls) phase in the hopping terms of the electrons'
Hamiltonian. In this paper we shall study a situation with a locally
staggered flux, created by plaquettes of ordered spins, where the
{\it global} flux is zero. This case is realized, for instance, in a Kagome
lattice (Ohgushi, Murakami and Nagaosa, 2000), a two dimensional (2D) lattice consisting of triangles
and hexagons of the same interatomic distance and that can be viewed as a
triangular Bravais lattice with a three-point atomic basis forming an
equilateral triangle of size half that of the triangular lattice parameter's.
In Fig.1 we remind the reader of the Kagome lattice structure.
                                                                                
\begin{figure}[h]
\begin{center}
\includegraphics[width=11cm]{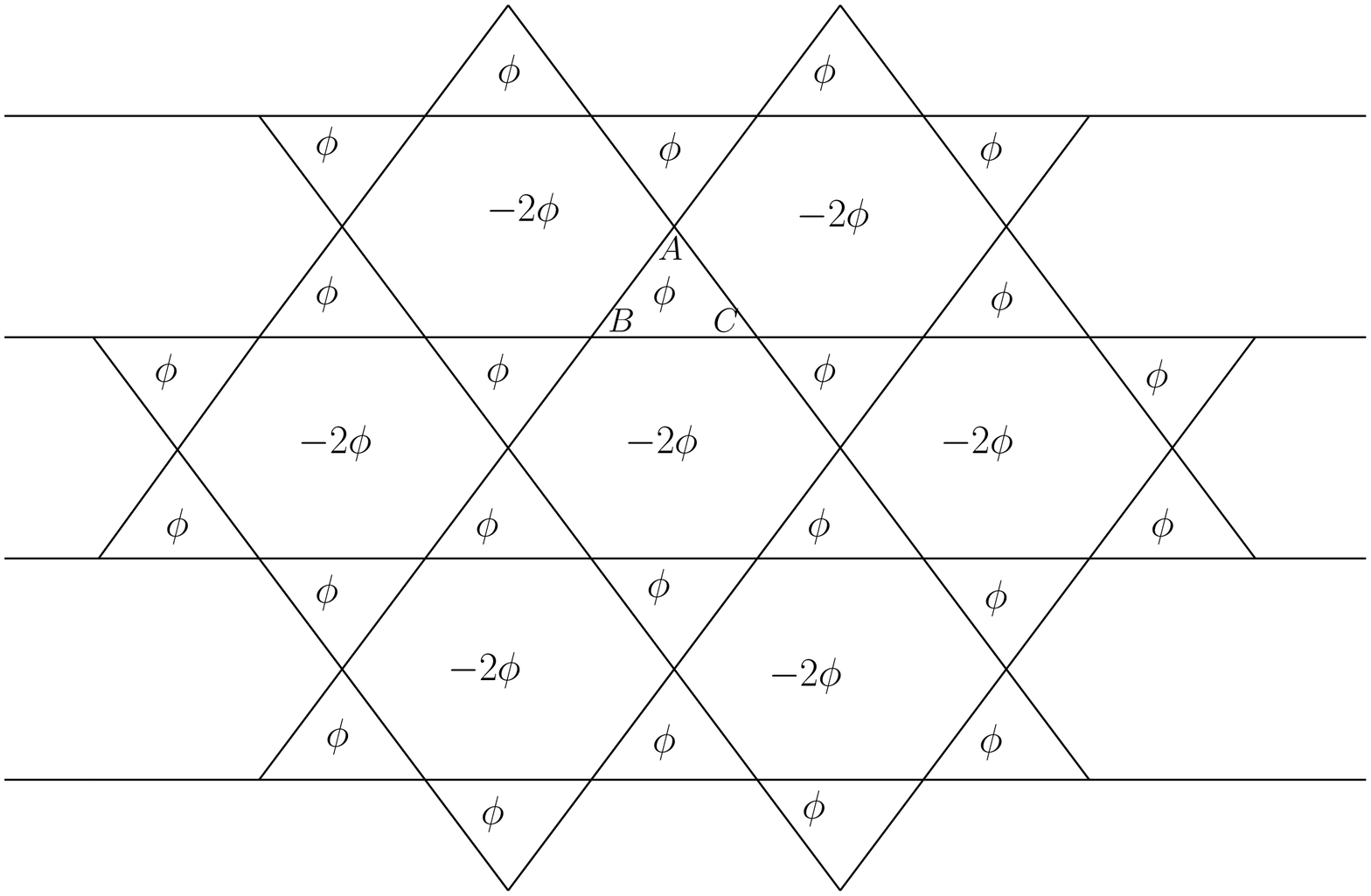}
\end{center}
\caption{Kagome lattice}
\end{figure}

\noindent
This type of lattice may have experimental relevance in the planes of 
pyrochlore compounds (Ohgushi, Murakami and Nagaosa, 2000, Ramirez, 1994, Harris and Zinkin, 1996) and the transport properties we 
describe may be appropriate for such materials. Ferromagnetic pyrochlore 
crystals of the type R$_2$Mo$_2$O$_7$ (R=Nd, Sm or Gd) have revealed 
interesting transport properties like an anomalous Hall effect increasing as 
the temperature $T$ is lowered, (Taguchi and Tokura, 1999) a feature that seems to be
connected with the geometrical frustration of pyrochlore lattices that is
partly embodied by the Kagome lattice itself, viewed now as the (1,1,1)
cross-section of the 3D pyrochlore's lattice. Motivated by the transport
properties of pyrochlore compounds, as well as by those of the manganite
ones, we study in this paper some quantum transport properties of the Kagome
lattice with a canted localised spin texture in which independent electrons
can move. 

The behavior of the Berry phase as a function of the spin canting for the
model at hand, and its consequences in terms of the macroscopic transport
properties of the model are studied in this paper. We discuss in detail the
energy spectrum as it depends strongly on the canting of the localized spins.
In particular, the nodes in the spectrum and the opening of energy gaps are
investigated, including their consequences for the transport properties.
Moreover, we evaluate explicitely the longitudinal conductivity $\sigma_{xx}$ 
and the Hall conductivity $\sigma_{xy}$ as a function of the canting angle 
$\theta$ or flux $\phi$. Our results for the transport properties are then 
compared with those of another famous model of this class, where a staggered 
magnetic field is applied to electrons within a honeycomb lattice (Haldane, 1988). 
The latter has very similar spectral properties as the model, first proposed 
by Ohgushi, Murakami and Nagaosa, (2000) defined on the Kagome lattice 
and studied here in greater detail. The model on the honeycomb lattice was 
proposed by Haldane as the condensed-matter (solid-state) equivalent of the 
quantum Hall effect, in that a quantization of the Hall $\sigma_{xy}$ 
conductivity can be achieved by varying the local flux per plaquette $\phi$, 
but without the need to introduce an external, homogeneous magnetic field. In 
the case of the Kagome lattice model at hand, the same result will be shown to
be attained through the introduction of a localised spin texture. Our
results confirm and complete the work by Ohgushi, Murakami and Nagaosa for
the Hall $\sigma_{xy}$ conductivity, with explicit calculations in terms of
expansions around the gap's nodes shown in detail, and moreover the
quantized values of the longitudinal $\sigma_{xx}$ conductivity are obtained
indicating the existence of some sort of metal-insulator transition as
the canting angle moves away from some special values.

The paper is organized as follows:
In Section 2 the tight-binding model of Ohgushi, Murakami and Nagaosa for
localized spins is described. The relationship between the localized spin's
wavefunction and the Berry phase of the hopping elctron is discussed in
Section 2.1, and the application to the Kagome lattice (Section 2.2) is
presented. In Section 2.3 the tigh-binding model for the honeycomb lattice
with a staggered magnetic field is also presented and compared to the model
on the Kagome lattice. Transport properties are then studied in Section 3
by making explicit use of Kubo's formula and an expansion of the energy 
spectrum near the nodes next to the Fermi energy, and the results obtained 
are discussed in Section 4.

\section{ A Model of Hopping Electrons in a Spin Texture } 

\subsection{ The Model } 

We consider the electronic hopping between nearest neighbours on a Kagome 
lattice as described by the tight-binding Hamiltonian. The electronic degrees 
of freedom are coupled to a set of localized spin-$S$ degrees of freedom on 
the same lattice via a local Hund coupling $J_H$. In this work $S=\frac{1}{2}$,
but generalization to the physically and theoretically interesting case of 
larger $S$ is possible. When $J_H$ is strong enough the spin of the hopping 
electron is forced to allign parallel to the localised spin ${\bf S}_i$ at 
each site and through double-exchange mechanism (Zener, 1951, Anderson and Hasegawa, 1955, de Gennes, 1960 ) the tight-binding 
hopping parameter $t_{ij}$ between two neighbouring sites $\langle i,j 
\rangle$ 
becomes proportional to the projection of the localised-spin wave function at 
site $j$ onto that at site $i$. The effective Hamiltonian representing the 
hopping is then
\begin{equation}
H=\sum_{\langle i,j \rangle}t_{ij}^{eff}c_{i}^{\dagger}c_{j} + {\rm h.c.}
\label{tbH}
\end{equation}
where $t_{ij}^{eff}=t\langle {\bf n}_i \vert {\bf n}_j \rangle$, $t$ 
being the bare hopping parameter and $\vert {\bf n} \rangle$ the spin wave 
function for a spin-$\frac{1}{2}$ quantized along the direction defined
by the unit vector 
${\bf n}=(\sin\theta\cos\phi,\sin\theta\sin\phi,\cos\theta)$. This spinor
wave function clearly satisfies (with 
$\vec{\sigma}=(\sigma_x,\sigma_y,\sigma_z)$ the vector of Pauli matrices) 
${\bf n}\cdot\vec{\sigma}\vert {\bf n} \rangle=+\vert {\bf n} \rangle$ and 
is given by
\begin{equation}
\vert {\bf n} \rangle = e^{ib} \left (
\begin{array}{c} 
\cos\frac{\theta}{2} \\
e^{i\phi}\sin\frac{\theta}{2}
\end{array}
\right ),
\end{equation}
where $b$ is an undetermined overall gauge degree of freedom. The effective 
hopping parameter is then
\begin{equation}
t_{ij}^{eff}=te^{-i(b_i-b_j)} \left \{ 
\cos\frac{\theta_i}{2}\cos\frac{\theta_j}{2}
+e^{-i(\phi_i-\phi_j)}\sin\frac{\theta_i}{2}\sin\frac{\theta_j}{2} \right \}
\end{equation}
and since $\vert \langle {\bf n}_i \vert {\bf n}_j \rangle \vert^2=
\cos^2\frac{\theta_{ij}}{2}$, with $\cos\theta_{ij}={\bf n}_i\cdot{\bf n}_j$ 
or $\theta_{ij}$ being the angle between the two localized spins' directions
of quantization so that
$\cos^2\frac{\theta_{ij}}{2}=\frac{1}{2}(1+{\bf n}_i\cdot{\bf n}_j)$, we
see that we can put 
\begin{equation}
t^{eff}_{ij}=te^{ia_{ij}}\cos\frac{\theta_{ij}}{2} 
\end{equation}
where the Berry phase $a_{ij}$ is obtained (ignoring the gauge parameters) 
through
\begin{equation}
e^{ia_{ij}}=\frac{ \cos\frac{\theta_i}{2}\cos\frac{\theta_j}{2}
+e^{-i(\phi_i-\phi_j)}\sin\frac{\theta_i}{2}\sin\frac{\theta_j}{2} }
{ \cos\frac{\theta_{ij}}{2} }
\end{equation}
and can be evaluated, e.g., by means of 
\begin{equation}
\sin a_{ij}=-\frac{ \sin\frac{\theta_i}{2}\sin\frac{\theta_j}{2}
\sin(\phi_i-\phi_j) }{ \cos\frac{\theta_{ij}}{2} }.
\end{equation}
To see what the phase $a_{ij}$ is, geometrically, we introduce the unit 
vector $\hat{z}$ and evaluate the triple product
\begin{equation}
{\bf n}_i\times{\bf n}_j\cdot\hat{z}=\sin\theta_i\sin\theta_j
\sin(\phi_i-\phi_j)=4\sin\frac{\theta_i}{2}\sin\frac{\theta_j}{2}
\cos\frac{\theta_i}{2}\cos\frac{\theta_j}{2}\sin(\phi_i-\phi_j)
\end{equation}
which shows that
\begin{equation}
\sin a_{ij}=-\frac{ {\bf n}_i\times{\bf n}_j\cdot\hat{z} }
{ 4\cos\frac{\theta_i}{2}\cos\frac{\theta_j}{2}\cos\frac{\theta_{ij}}{2} }.
\end{equation}
This expression is a special case of the formula giving the solid angle 
$\Omega({\bf n}_1,{\bf n}_2,{\bf n}_3)$ between three unit vectors 
${\bf n}_1$, ${\bf n}_2$ and ${\bf n}_3$:
\begin{equation}
\sin\frac{\Omega({\bf n}_1,{\bf n}_2,{\bf n}_3)}{2}=
\frac{ {\bf n}_1\times{\bf n}_2\cdot{\bf n}_3 }
{ 4\cos\frac{\theta_{12}}{2}\cos\frac{\theta_{13}}{2}
\cos\frac{\theta_{23}}{2} }
\end{equation}
(as can be verified by taking, e.g., ${\bf n}_1=\hat{x}$, ${\bf n}_2=\hat{y}$
and ${\bf n}_3=\hat{z}$; $\Omega({\bf n}_1,{\bf n}_2,{\bf n}_3)$ can also be
seen as the area of the portion of unit sphere enclosed by the maximum 
circles passing through the unit vectors' tips). In the last formula, of 
course, $\cos\theta_{kk'}={\bf n}_k\cdot{\bf n}_{k'}$ and therefore 
$\cos^2\frac{\theta_{kk'}}{2}=\frac{1}{2} (1+{\bf n}_k\cdot{\bf n}_{k'})$. 
We remark that this formula for three spins is completely analogous to that 
for the chirality gauge field in the formulation of Lee and Nagaosa 
(1992) for the chiral spin liquid theory of high-temperature 
superconductivity (Wen, Wilczek and Zee, 1989). In this formulation the instantaneous gauge 
field flux through the triangular plaquette made up by the three spins is 
$\Phi({\bf n}_1,{\bf n}_2,{\bf n}_3)=
\frac{1}{2} \Omega({\bf n}_1,{\bf n}_2,{\bf n}_3)$.  Back to our two-spins
hopping problem, we then conclude that 
$\sin  a_{ij}=-\sin\frac{1}{2}\Omega({\bf n}_i,{\bf n}_j,\hat{z})$, or
\begin{equation}
a_{ij}=\pi+\frac{1}{2} \Omega({\bf n}_i,{\bf n}_j,\hat{z})
\end{equation}
(the factor $\frac{1}{2}$ being probably due to our specially chosen 
localised spin value, which leads to the conjecture that for a generic 
spin-$S$ situation the Berry phase would be 
$a_{ij}=\pi+S\Omega({\bf n}_i,{\bf n}_j,\hat{z})$). Since the solid angle 
$\Omega({\bf n}_i,{\bf n}_j,\hat{z})$ is also the unit sphere's surface 
area between the tips of the three vectors ${\bf n}_i$, ${\bf n}_j$ and 
$\hat{z}$, the phase $a_{ij}=\int_i^jd{\bf r}\cdot{\bf A}$ can also be seen 
as the flux of a magnetic monopole's field of modulus 
$\vert {\bf B} \vert=\frac{1}{2}$ with the monopole placed in the sphere's 
center, or, alternatively, as the flux of the related gauge field ${\bf A}$ 
through the triangle bearing on the segment $(i,j)$ of a triangular Kagome 
lattice's unit cell (Ohgushi, Murakami and Nagaosa, 2000). In this way, the Berry phase $a_{ij}$ 
acquires some physical meaning too. 

We now consider this tight-binding model on the Kagome lattice with a 
fixed localised-spin configuration (or spin texture) as suggested by Ohgushi, Murakami and Nagaosa (2000) 
, in which the unit vectors ${\bf n}_i$ at each site of a 
triangular unit cell are tilted outwards at a fixed angle $\theta$ over the 
unit vector $\hat{z}$ orthogonal to the lattice plane. This means (labelling 
the spins clockwise A, B and C in the unit cell)
\begin{equation}
\sin a_{AB}=\sin a_{BC}=\sin a_{CA}=\frac{\sqrt{3}\sin^2\theta}
{4(1+\cos\theta)\sqrt{1-\frac{3}{4}\sin^2\theta}}.
\end{equation}  
The flux generated by the spins in every triangular unit is set equal to 
$\phi$ with the condition
\begin{equation}
e^{i\phi}=e^{i(a_{AB}+a_{BC}+a_{CA})}=e^{3ia_{AB}}
\end{equation}
with $\phi=3a_{AB}$ (mod $2\pi$) and thus
\begin{equation} 
\sin\frac{\phi}{3}=\frac{\sqrt{3}(1-\cos\theta)}
{2\sqrt{1+3\cos^2\theta}}.
\end{equation}
This is equivalent to the expression proposed by Ohgushi, Murakami and
Nagaosa (preprint of Ohgushi, Murakami and Nagaosa (2000)) $\phi=\pi+3\arg(1-i\sqrt{3}\cos\theta)$. 
The graph for this expression of $\phi=\phi(\theta)$ is shown in Fig.2 for 
convenience.

\begin{figure}[h]
\begin{center}
\includegraphics[width=11cm]{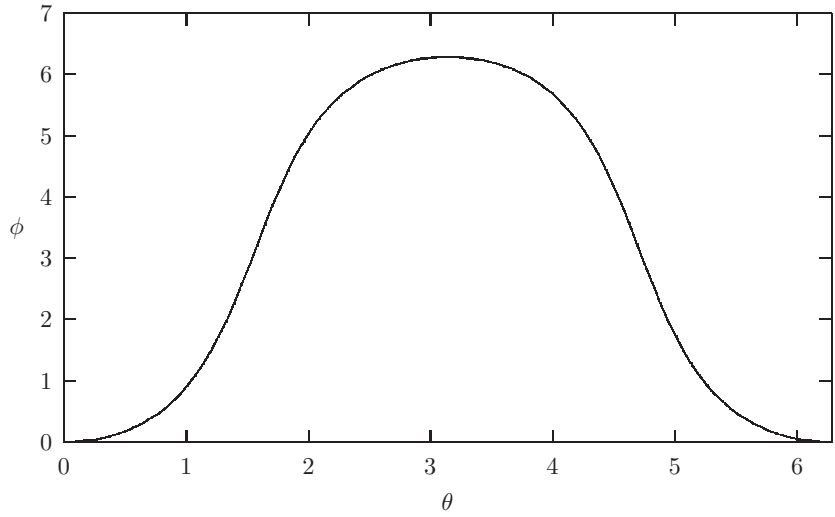}
\end{center}
\caption{Flux $\phi$ as a function of the tilting angle $\theta$
of the localized spins.}
\end{figure}

As pointed out by Ohgushi, Murakami and Nagaosa (2000), the flux per triangular unit cell $\phi$ is 
cancelled out for the chosen spin texture by the flux $-2\phi$ generated by 
each of the remaining hexagonal hopping plaquettes on the Kagome lattice. 
There are indeed twice as many triangular units as hexagonal plaquettes, so 
that the overall gauge field flux through the lattice is zero. This situation 
is reminiscent of the analogous tight-binding model in a staggered magnetic 
field as was proposed by Haldane (1988) to mimick the quantized Hall 
effect in a condensed-matter situation. In the present model, the chosen 
spin-texture, with all localised spins tilted by the same angle $\theta$ in 
each unit cell, is presumably the one corresponding to the mean-field 
solution for some magnetic spin-spin Heisenberg Hamiltonian which should be 
added to our tight-binding Hamiltonian, Eq. (\ref{tbH}), to give a total 
Hamiltonian of the type
\begin{equation}
H_{tot}=\sum_{\langle i,j \rangle}t_{ij}^{eff}(\{{\bf S}_i^{(0)}\})
c_{i}^{\dagger}c_{j} + {\rm h.c.} + \sum_{i,j}J_{ij}{\bf S}_i\cdot{\bf S}_j.
\end{equation}
The role of the spin fluctuations around this ordered spin texture, 
$\{{\bf S}_i^{(0)}\}$, as well as the effects of different spin textures, 
(e.g. AFM ones) could serve as an interesting further research problem for
future studies.   

\subsection{ Band structure for the Kagome lattice } 

The Kagome lattice is made up of triangular and hexagonal plaquettes and can
be seen as a triangular lattice with a 3-point basis where every triangular 
unit cell contains three sites, $A$, $B$, $C$ (Fig 1). The displacement 
vectors between these sites are $\vec{a}_{1}=(-1/2,-\sqrt{3}/2)$, 
$\vec{a}_{2}=(1,0)$ and $\vec{a}_{3}=(-1/2,\sqrt{3}/2)$, with 
$\sum_i \vec{a}_i=0$. The effective hopping parameter of (\ref{tbH}) can be 
written as 
\begin{equation}
t^{eff}_{ij}=te^{ia_{ij}}\cos(\frac{\theta_{ij}}{2})
\end{equation}
and since $\cos(\theta_{ij}/2)=\sqrt{1-\frac{3}{4}\sin^2\theta}$ is fixed 
for the chosen spin texture, $\theta_{ij}$ being the angle between the n.n. 
pair of localized spins, we can choose the convention where 
$t\cos(\frac{\theta_{ij}}{2})\equiv 1$. Then, in momentum space, the 
Hamiltonian can be rewritten as
\begin{equation}
H=\sum_{\vec{k}}\psi^{\dagger}(\vec{k})h(\vec{k})\psi(\vec{k})
\end{equation}
where $\psi(\vec{k})=\left( c_A(\vec{k}),c_B(\vec{k}),c_C(\vec{k}) \right)$ 
and $h(\vec{k})$ is the (suitably symmetrized) matrix
\begin{eqnarray}
h(\vec{k})=\left (
\begin{array}{ccc}
0 & 2\cos(\vec{k}\cdot\vec{a}_{1})e^{-i\phi/3} &  
2\cos(\vec{k}\cdot\vec{a}_{3})e^{i\phi/3}\\
2\cos(\vec{k}\cdot\vec{a}_{1})e^{i\phi/3} & 0 & 
2\cos(\vec{k}\cdot\vec{a}_{2})e^{-i\phi/3}\\
2\cos(\vec{k}\cdot\vec{a}_{3})e^{-i\phi/3} &  
2\cos(\vec{k}\cdot\vec{a}_{2})e^{i\phi/3} & 0
\end{array}
\right ).
\end{eqnarray}
The three eigenvalues of this Hamiltonian are
\begin{eqnarray}
E_{up}(\vec{k})&=&4\sqrt{\frac{1+f(\vec{k})}{3}}\cos(\frac{\theta(\vec{k})}{3})
\nonumber\\
E_{mid}(\vec{k})&=&4\sqrt{\frac{1+f(\vec{k})}{3}}\cos(\frac{\theta(\vec{k})
-2\pi}{3})\\
E_{down}(\vec{k})&=&4\sqrt{\frac{1+f(\vec{k})}{3}}\cos(\frac{\theta(\vec{k})
+2\pi}{3})\nonumber
\end{eqnarray}
with
\begin{equation}
\theta(\vec{k})=arg \left[ f(\vec{k})\cos(\phi)+i\sqrt{4\left 
(\frac{1+f(\vec{k})}{3}\right )^{3}-(f(\vec{k})\cos(\phi))^{2}} \right]
\end{equation}
and
\begin{equation}
f(\vec{k})=2\cos(\vec{k}\cdot\vec{a}_{1})\cos(\vec{k}\cdot\vec{a}_{2})
\cos(\vec{k}\cdot\vec{a}_{3}).
\end{equation}
The three bands touch in six points only for $\phi=0$ and $\phi=\pm \pi$, 
while for $\phi$ different from these values there is a gap between the bands,
as shown in Fig. 3 ($\phi=0$) and Fig. 4 ($\phi=\pi/3$).
\\
\\
\begin{figure}[h]
\begin{center}
\includegraphics[width=9cm]{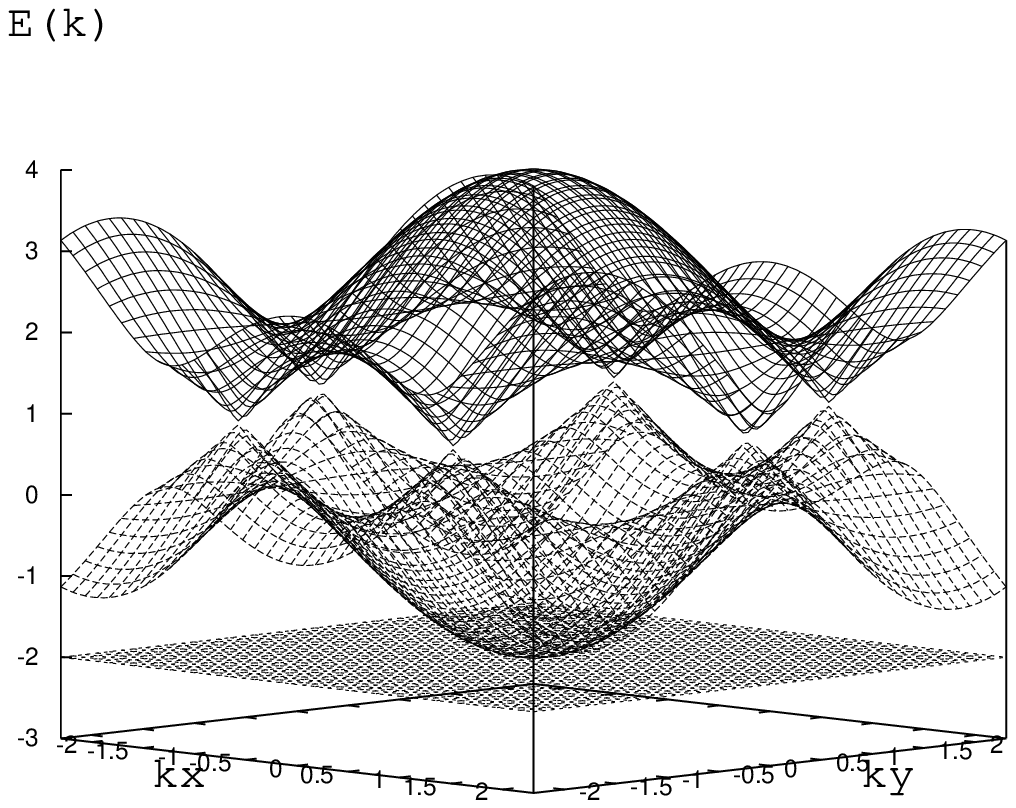}
\end{center}
\caption{Eigenvalues for the Kagome lattice for $\phi=0$}
\end{figure}

\begin{figure}[h]
\begin{center}
\includegraphics[width=9cm]{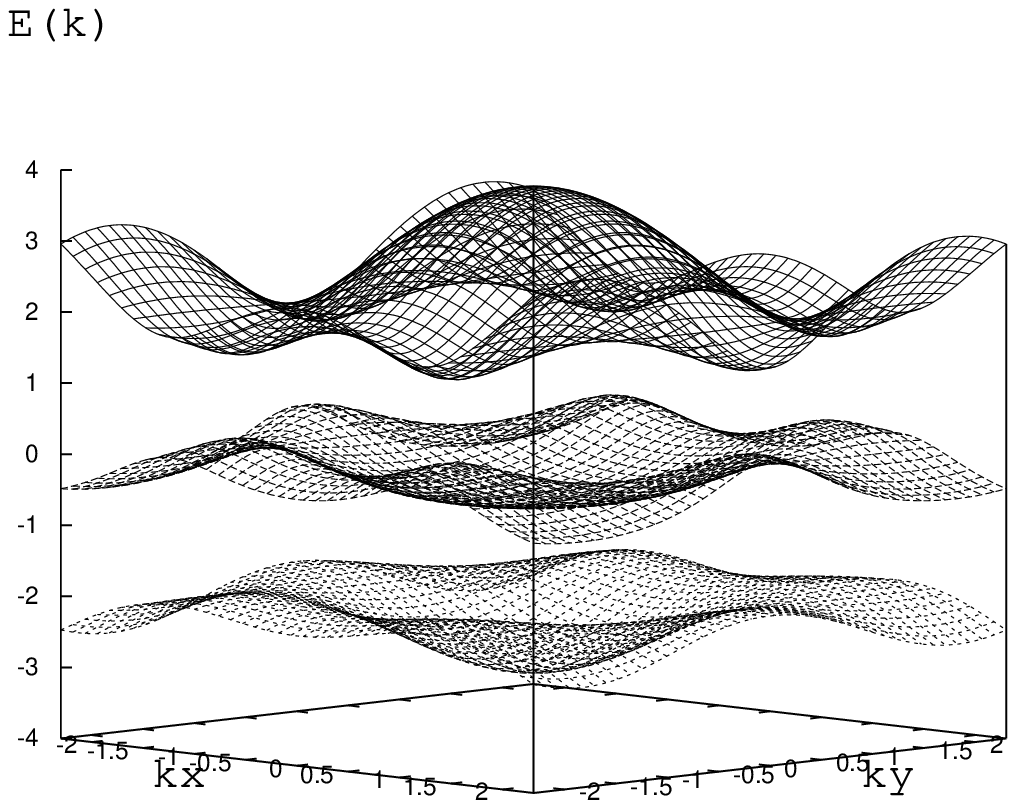}
\end{center}
\caption{Eigenvalues for the Kagome lattice for $\phi=\pi/3$}
\end{figure}

\noindent
These nodes are: $(\pm\frac{2\pi}{3},0)$ and
$(\pm\frac{\pi}{3},\pm\frac{\sqrt{3}}{3}\pi)$, on the vertices of a hexagon,
as shown in Fig. 5.

\begin{figure}[h]
\begin{center}
\includegraphics[width=13cm]{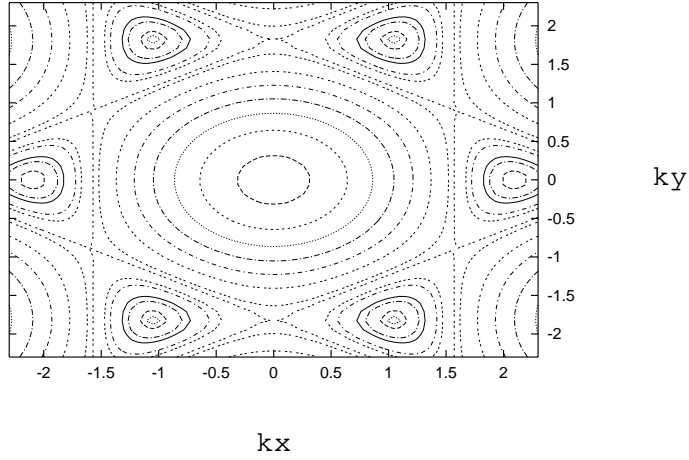}
\end{center}
\caption{Nodes position for the Kagome lattice}
\end{figure}

The problem of calculating transport coefficients with this $3 \times 3$ 
matrix is not exactly solvable, so we reduce this matrix to a $2\times 2$ one
by expanding $h(\vec{k})$ around the nodes. This can be done with a unitary 
transformation which allows us to neglect the terms related to the lower band; 
in fact this band is far from the other two and gives no relevant contribution 
to the Green function present in the Kubo formula. To find this unitary 
transformation we consider the Hamiltonian evaluated at a node ${\bf k}_0$. 
If we apply the unitary matrix that diagonalizes $h(k_{x0},k_{y0})$, where 
$(k_{x0},k_{y0})$ are the node's coordinates, to the Hamiltonian evaluated 
at the general point $(k_{x},k_{y})$, we find a matrix $H'$ with the 
structure
\begin{eqnarray}
H'=
\left(
\begin{array}{ccc}
\lambda_{1} & \alpha[\tilde{k}_{x}-i\tilde{k}_{y}] & 
\beta[\tilde{k}_{x}-i\tilde{k}_{y}]\\
\bar{\alpha}[\tilde{k}_{x}+i\tilde{k}_{y}] &\lambda_{2} & 
\gamma[\tilde{k}_{x}+i\tilde{k}_{y}]\\
\bar{\beta}[\tilde{k}_{x}+i\tilde{k}_{y}] &
\bar{\gamma}[\tilde{k}_{x}-i\tilde{k}_{y}] & \lambda_{3}
\end{array}
\right)
\end{eqnarray}
with $\tilde{k}_{x}=(k_{x}-k_{x0})$ and $\tilde{k}_{y}=(k_{y}-k_{y0})$.
The elements on the diagonal are the eigenvalues of the Hamiltonian
$h(k_{x0},k_{y0})$, while the off-diagonal elements are complex combinations 
of ${\tilde k}_{x,y}$. Near the nodes the distance between the 
upper and the middle band is small, while the lower band is distant and 
gives no relevant contribution. To justify this we can consider a 
projection of the Green function around the node. \\
We want to find a projection only on the first two eigenvalues, so we 
choose a projector such that
\begin{equation}
PH^{'}P = \left(
\begin{array}{cc}
\lambda_{1} & H^{'}_{12}\\
H^{'}_{21} & \lambda_{2}
\end{array}
\right).
\end{equation}
We define the Green function as $G=(z-H^{'})^{-1}$ and the projection 
operator $P$ with the convention that $(A)^{-1}_{P}=(PAP)^{-1}$ is 
the inverse operation on the projected space, $(1-P)$ being the 
projection operator complementary to $P$. The Green function can now be 
written as
\begin{equation}
G=PGP+(1-P)GP+PG(1-P)+(1-P)G(1-P)
\end{equation}
and the projected Green function is 
\begin{eqnarray}
PGP=P(z-H^{'})^{-1}P=(z-PH^{'}P-PH^{'}(1-P)
(z-H^{'})^{-1}_{1-P}(1-P)H^{'}P)^{-1}_{P}. 
\nonumber
\end{eqnarray}
The last terms are of higher order and can be neglected. If we consider the 
terms related to the lower eigenvalue, we can see that 
$1/(z-\lambda_{3})\sim 1/2$, because we are considering $\vec{k}$ near the 
nodes. This eigenvalue does not give an important contribution, so it can be 
neglected and we can use the approximation
\begin{equation}
PGP\approx(z-PH^{'}P)^{-1}_{P} .
\end{equation}
Now we can write the projection of the Hamiltonian $H'$ as
\begin{eqnarray}
PH'P&=&\left(
\begin{array}{cc}
\lambda_{1} & \alpha[\tilde{k}_{x}-i\tilde{k}_{y}]\\
\bar{\alpha}[\tilde{k}_{x}+i\tilde{k}_{y}] & \lambda_{2}
\end{array}
\right)\nonumber\\
\nonumber\\
&=&\left(
\begin{array}{cc}
1+\frac{\sqrt{3}\phi}{3} & h_{1}-ih_{2}\\
h_{1}+ih_{2} & 1-\frac{\sqrt{3}\phi}{3}
\end{array}
\right)\\
\nonumber\\
&=&I+\left(
\begin{array}{cc}
 m & h_{1}-ih_{2}\\
 h_{1}+ih_{2} & -m
\end{array}
\right)=I+h\nonumber
\end{eqnarray}
where $I$ is the identity matrix, with 
\begin{eqnarray}
h_{1}&=&\alpha_{1}\tilde{k}_{x}+\alpha_{2}\tilde{k}_{y}\nonumber\\
h_{2}&=&-\alpha_{2}\tilde{k}_{x}+\alpha_{1}\tilde{k}_{y}\\
m&=&\frac{\sqrt{3}}{3}\phi ,\nonumber
\end{eqnarray}
with $\alpha_{1}$ and $\alpha_{2}$ are the components of a complex parameter, 
depending on the node we are considering. The new matrix representing the 
Hamiltonian has eigenvalues $\pm \lambda$ and eigenvectors $\Psi_{\pm}$ with
\begin{eqnarray}
\lambda&=&\sqrt{m^{2}+h_{1}^{2}+h_{2}^{2}}\nonumber\\
\Psi_{\pm}&=&\frac{1}{\sqrt{1+(\pm\lambda-m)^{2}/|k|^{2}}}\left(
\begin{array}{c}
1\\
\frac{\pm\lambda-m}{k}
\end{array}
\right).\nonumber
\end{eqnarray}
As was said, the bands touch only when $\phi=0$ and $\phi=\pm\pi$. Here we 
consider $\phi$ different from zero, but small enough so that we create a small 
gap between the two bands.

\subsection{ Band structure of the honeycomb lattice } 

Here we also consider the case of a honeycomb lattice, as was first envisaged by 
Haldane (1988); this is made up by two sublattices that we call A and B 
(Fig. 6), or by a triangular lattice with a 2-point basis. The displacement 
vectors from a B site to the three nearest neighbours are: 
$\vec{a}_{1}=(-\sqrt{3}/2,1/2)$, $\vec{a}_{2}=(0,-1)$ and 
$\vec{a}_{3}=(\sqrt{3}/2,1/2)$, while the displacement vectors from the site 
B and the nearest neighbours on the same sublattice are
$\vec{b}_{1}=(-\sqrt{3}/2,-3/2)$, $\vec{b}_{2}=(\sqrt{3},0)$ and  
$\vec{b}_{3}=(-\sqrt{3}/2,3/2)$ (again, $\sum \vec{b}_i=0$). 

\begin{figure}[h]
\begin{center}
\includegraphics[width=9cm]{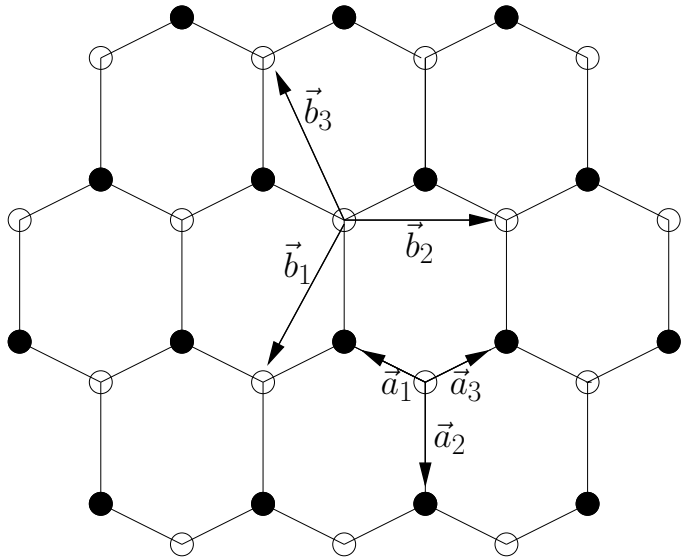}
\end{center}
\caption{Honeycomb lattice}
\end{figure}

Here too, we consider a tight-binding model in the presence of a staggered 
magnetic flux (Haldane, 1988); the Hamiltonian for this Haldane model is
$H=\sum_{\vec{k}}\psi^{\dagger}(\vec{k})h(\vec{k})\psi(\vec{k})$ with

\begin{eqnarray}
h(\vec{k})=&&2t_{2}\cos{\phi}\sum_{i}\cos({\bf k\cdot b_{i}}) 
{\bf I}+t_{1}\sum_{i}[\cos({\bf k\cdot a_{i}}){\bf \sigma_1}
+\sin({\bf k\cdot a_{i}}){\bf\sigma_2}]\nonumber\\
+&&[M-2t_{2}\sin{\phi}\sum_{i}\sin({\bf k\cdot b_{i}})]{\bf \sigma_3},
\end{eqnarray}
where $t_{1}$ is a hopping parameter between nearest neighbours on different 
sublattices, $t_{2}$ is a hopping parameter between nearest neighbour sites 
on the same sublattice, and $\sigma_i$ are the three Pauli matrices.
If we rewrite the Hamiltonian as $H=a\sigma_1+b\sigma_2+c\sigma_3$, the 
Hamiltonian matrix reads
\begin{eqnarray}
H=\left (
\begin{array}{cc}
c & a-ib\\
a+ib & -c
\end{array}
\right ).
\end{eqnarray}

\begin{figure}[h]
\begin{center}
\includegraphics[width=9cm]{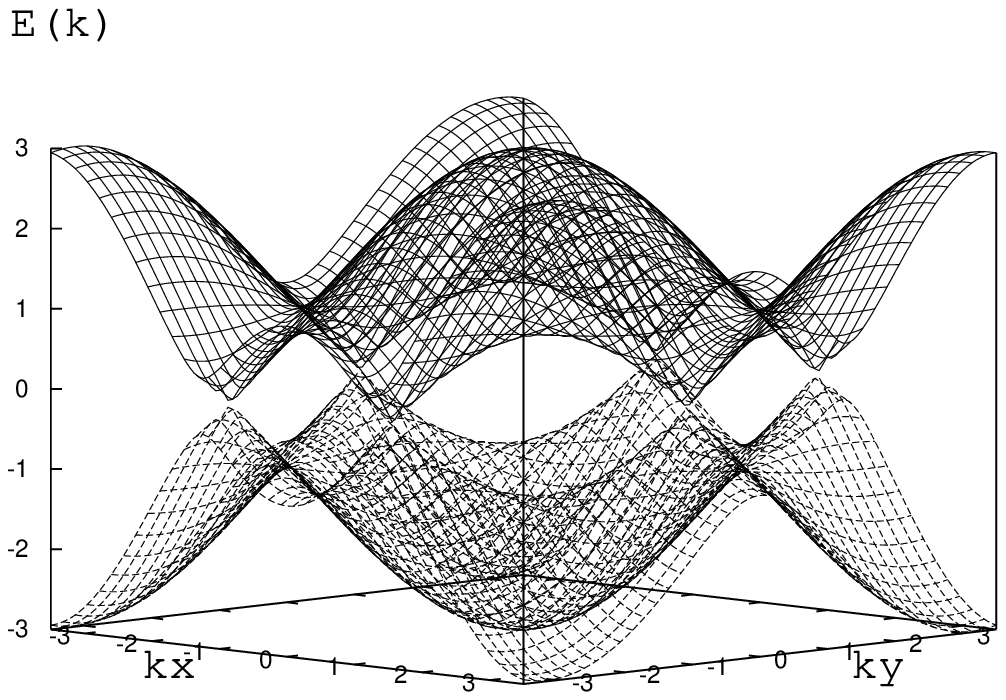}
\end{center}
\caption{Eigenvalues for the honeycomb lattice for $\phi=0$}
\end{figure}

\begin{figure}[h]
\begin{center}
\includegraphics[width=9cm]{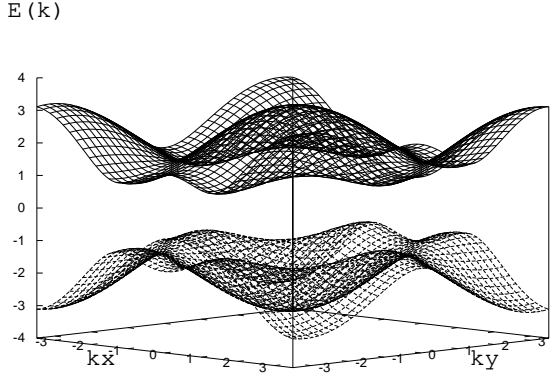}
\end{center}
\caption{Eigenvalues for the honeycomb lattice for $\phi=\frac{\pi}{3}$,
$M\neq0$ and $M\neq 3\sqrt{3}\alpha t_{2}\sin{\phi}$}
\end{figure}

The eigenvalues of this matrix are $\pm \lambda$ with 
$\lambda=\sqrt{a^{2}+b^{2}+c^{2}}$ (Fig. 7 ($\phi=0$) and Fig. 8 
($\phi=\frac{\pi}{3}$)), while the eigenvectors are \begin{equation}
\psi_{\pm}=\frac{1}{\sqrt{1+\frac{(\pm \lambda)^{2}}{|k|^{2}}}}\left (
\begin{array}{c}
1\\
\frac{\pm\lambda-c}{k}
\end{array}
\right )
\end{equation}
where $k=a-ib$. 
Formally this case is similar to that of the Kagome lattice, after the 
reduction of the original matrix to a $2 \times 2$ one. The two bands meet 
when the condition $a^{2}+b^{2}+c^{2}=0$ is satisfied. This becomes a 
condition on the parameter $M$: there are nodes when 
$M=3\sqrt{3}t_{2}\alpha\sin{\phi}$, with $\alpha=\pm 1$. When $\phi=0$ and 
$M=0$ there are six nodes: $(\pm\frac{4\pi}{3\sqrt{3}},0)$ and 
$(\pm\frac{2\pi}{3\sqrt{3}},\pm\frac{2\pi}{3})$ (Fig. 9), while when 
$M=3\sqrt{3}t_{2}\alpha\sin{\phi}$ there are only three of these nodes.

\begin{figure}[h]
\begin{center}
\includegraphics[width=13cm]{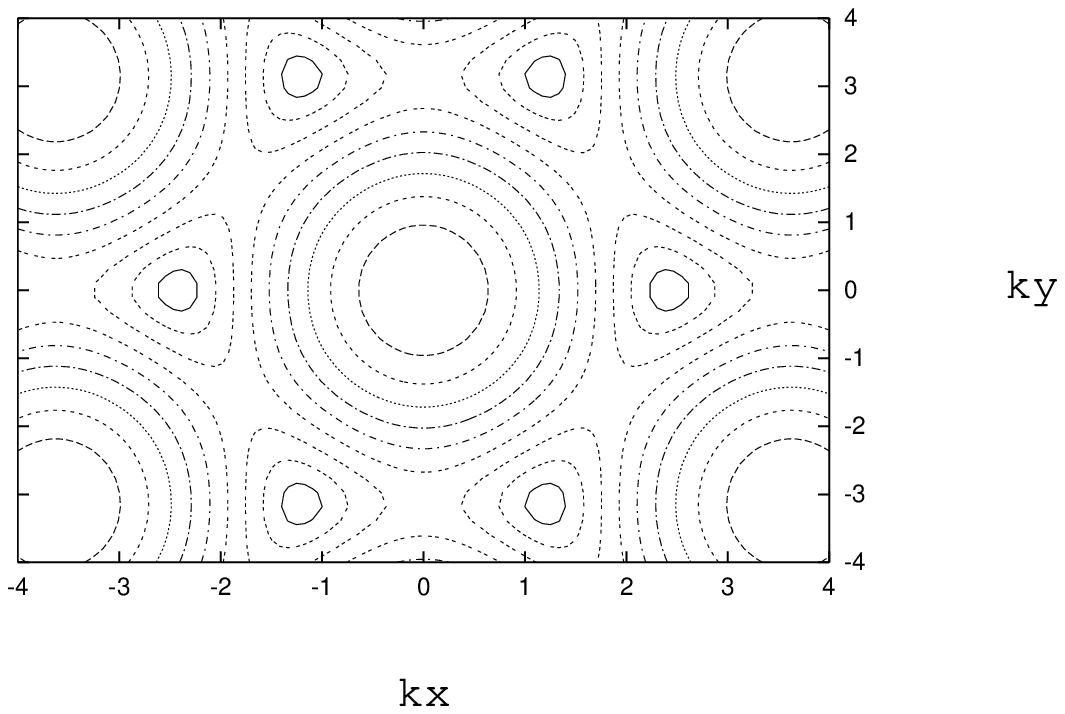}
\end{center}
\caption{Nodes position for the honeycomb lattice for $\phi=0$}
\end{figure}

\section{ Transport Properties}

Based on the linear-response theory, a suitable Kubo formula and the 
corresponding conductivity tensor can be studied for the Hamiltonians 
considered in Sect. 2. Some details are given in Appendix A. From this result 
we can derive the Hall conductivity $\sigma_{xy}$ and the longitudinal 
conductivity $\sigma_{xx}$ of our two-dimensional tight-binding model.
We verify explicitely the quantization of $\sigma_{xy}$ as a function of
$\phi$ (or $\theta$) and calculate explicitely the longitudinal conductivity
$\sigma_{xx}$ which also appears to be quantized in the absence of disorder
or other symmetry-breaking conditions. For the two models considered, these 
are our main new results.

\subsection{Hall conductivity } 

For $\mu\neq\nu$ the third term in Eq. (\ref{kubo}) (Appendix A)
vanishes and after the integration with respect to $E$ we 
find that the Hall conductivity, in the limit of $\omega=0$ and $T=0$, is
\begin{equation}
\sigma_{xy}=-\frac{1}{\hbar \eta}Re\sum_{k}~\lambda_{k}~
[U^{\dagger}j_{\mu}(h-\lambda_{k}+2i\eta)^{-1}j_{\nu}U]_{kk} ,
\end{equation}
where $U$ is the unitary matrix that diagonalizes the Hamiltonian matrix 
$h(\vec{k})$, while $j_{x}$ and $j_{y}$ are the current matrices. From this 
we find that the Hall conductivity for every node $n$ for the case of the 
Kagome lattice is
\begin{equation}
\sigma_{xy}^{n}=\frac{e^{2}}{\hbar\eta}
\int_{-\infty}^{+\infty}\frac{12m\eta}{8(m^{2}
+h_{1}^{2}+h_{2}^{2})^{\frac{3}{2}}}\frac{d^{2}k}{(2\pi)^{2}}
=\frac{e^{2}}{2h}sgn(\phi) .
\end{equation}
The integration being over the hexagonal Brillouin zone, we considered 
only one third of the integral and then we have to multiply for the number 
of nodes. We can conclude that the Hall conductivity is different from 
zero and it is quantized in the presence of a gap between the bands (that 
is for $\phi$ different from 0, $\pm \pi$) and is equal to
\begin{equation}
\sigma_{xy}=\frac{e^{2}}{h} sgn(\phi).
\end{equation}
So, we have another model of transverse conductivity quantization in the 
absence of an external uniform magnetic field.

Now we consider the case of the honeycomb lattice, in Haldane's model. For 
$M=0$ and $\phi=0$ the bands touch in six points: 
$(\pm \frac{4\pi}{3\sqrt{3}},0)$, 
$(\pm\frac{2\pi}{3\sqrt{3}},\pm\frac{2\pi}{3})$ and the Hamiltonian is 
simply of the form
\begin{eqnarray}
\left (
\begin{array}{cc}
0 & a-ib\\
a+ib & 0
\end{array}
\right ).
\end{eqnarray}
Expanding the terms around the nodes, we find that the function to integrate 
in order to find the Hall conductivity is
\begin{equation}
\pm\frac{9ab}{8(a^{2}+b^{2})^{\frac{3}{2}}}
\end{equation}
but the integral of this term gives zero contribution.
To generate a gap we have to move from the situation in which $M=0$ and 
$\phi=0$. We add a small mass contribution $M\ll 1$, but we mantain $\phi=0$;
now the Hamiltonian is
\begin{eqnarray}
\left (
\begin{array}{cc}
M & a-ib\\
a+ib & -M
\end{array}
\right ).
\end{eqnarray}
In this case the function to integrate is
\begin{equation}
\pm\frac{9t^{2}_{1}M\eta}{8(a^{2}+b^{2}+M^{2})^{\frac{3}{2}}};
\end{equation}
three nodes give a positive contribution and three a negative one, to give
\begin{equation}
\pm\frac{e^{2}}{2h}sgn(M)
\end{equation}
Summing up all the contributions we find that the Hall conductivity is zero. 
Different is the situation in which we consider $M=0$, but we add a small 
flux $\phi$. We rewrite the Hamiltonian as
\begin{eqnarray}
\left (
\begin{array}{cc}
c & a-ib\\
a+ib & -c
\end{array}
\right ),
\end{eqnarray}
where
\begin{equation}
c=-2t_{2}\sin\phi~\sum_{i}\sin({\bf k\cdot b_{i}}).
\end{equation}
Now the function to integrate is ($k=a-ib$)
\begin{equation}
P(k)=\pm \frac{9c\eta t^{2}_{1}}{8(a^{2}+b^{2}+c^{2})^{\frac{3}{2}}}.
\end{equation}
Near three of the six nodes $P(k)$ is negative, but approximating $c$ around 
these points we find $c\simeq 3\sqrt{3}t_{2}\sin\phi$, so the function to 
integrate is negative. After the integration we find that every one of these 
three points gives a conductivity equal to
\begin{equation}
\frac{e^{2}}{2h}sgn(\sin\phi).
\end{equation}
For the other three points $P(k)$ is positive, but the expansion of $c$ is 
$c\simeq -3\sqrt{3}t_{2}\sin(\phi)$. So, now the function to integrate 
is negative too and it gives the same result as before. Summing over all the 
points and remembering that the integration is over the exagon we find that 
the Hall conductivity for the honeycomb lattice case is
\begin{equation}
\sigma_{xy}=\frac{e^{2}}{h}sgn(\sin\phi).
\end{equation}
We can conclude that the Hall conductivity can be rewritten as
\begin{equation}
\sigma_{xy}=\nu~\frac{e^{2}}{h}
\end{equation}
with $\nu=\pm 1$, depending on the sign of $\phi$.

\subsection{ Longitudinal conductivity }

Here too we use the general expression derived from the Kubo formula. In this 
case ($\mu=\nu=x$) the longitudinal conductivity $\sigma_{xx}$ derived by 
expression (\ref{kubo}), after the energy integration, is
\begin{eqnarray}
-\frac{1}{\hbar}\sum_{k,m}\frac{\rho_{0}(\lambda_{k})(U^{\dagger} j_{x}U)_{km}
(U^{\dagger} j_{x}U)_{mk}}{-\omega+2i\eta}
\Big[\frac{1}{\lambda_{m}-\lambda_{k}-\omega+2i\eta}
+\frac{1}{\lambda_{m}-\lambda_{k}+\omega-2i\eta} \Big].\nonumber
\end{eqnarray}
After summing over the eigenvalues, substituting the values of $\lambda_k$ 
and $\epsilon=\eta+\frac{i\omega}{2}$ we find
\begin{equation}
\sigma_{xx}=-\frac{1}{2\hbar\epsilon}\int_{-\infty}^{+\infty}
(U^{\dagger}j_{x}U)_{21}(U^{\dagger}j_{x}U)_{12}
\frac{\sqrt{m^{2}+h_{1}^{2}+h_{2}^{2}}}{m^{2}+h_{1}^{2}+h_{2}^{2}
+\epsilon^{2}}\frac{d^{2}k}{(2\pi)^{2}}.
\label{sigmaxx}
\end{equation}
>From this expression, we have still to subtract the diamagnetic term and so
we evaluate
(Ludwig, Fisher, Shankar and Grinstein, 1994)
\begin{equation}
\tilde{\sigma}_{xx}=\sigma_{xx}-\frac{1}{\epsilon}
\lim_{\epsilon\rightarrow0}\epsilon\sigma_{xx}.
\label{subtr}
\end{equation}
For the Kagome lattice the product of the matrix elements of the currents 
is
\begin{equation}
(U^{\dagger}j_{x}U)_{21}(U^{\dagger}j_{x}U)_{12}=\frac{3(4m^{2}+3h_{1}^{2}
+2\sqrt{3}h_{1}h_{2}+h_{2}^{2})}{4(m^{2}+h_{1}^{2}+h_{2}^{2})},
\end{equation}
and remembering that $h_{1}$ and $h_{2}$ are symmetric variables and using 
polar coordinates we can rewrite $\sigma_{xx}$ as 
\begin{eqnarray}
\sigma_{xx}=-\frac{e^{2}}{4h\epsilon}\Big[\int_{0}^{\infty}
\frac{m^{2}r}{\sqrt{m^{2}+r^{2}}(m^{2}+r^{2}+\epsilon^{2})}dr
+\int_{0}^{\infty}\frac{r\sqrt{m^{2}+r^{2}}}{m^{2}+r^{2}+\epsilon^{2}}dr
\Big],
\end{eqnarray}
so that carrying out the integrals we find
\begin{eqnarray}
\int_{0}^{\infty}\frac{rm^{2}}{\epsilon\sqrt{m^{2}+r^{2}}(m^{2}+r^{2}
+\epsilon^{2})}dr=
\left\{
\begin{array}{lr}
\frac{m^{2}}{\epsilon^{2}}\arccos(\frac{m}{\sqrt{m^{2}
+\epsilon^{2}}}) & \mbox{if } m>0 \\
0 & \mbox{if } m=0 \\
\frac{m^{2}}{\epsilon^{2}}(\pi-\arccos(\frac{m}{\sqrt{m^{2}+\epsilon^{2}}}) 
& \mbox{if } m<0
\end{array}
\right.
\end{eqnarray}
For the second integral we introduce a cut-off $\lambda$ and evaluate
\begin{eqnarray}
\int_{0}^{\lambda}\frac{r\sqrt{m^{2}+r^{2}}}{\epsilon(m^{2}+r^{2}
+\epsilon^{2})}dr=
\left\{
\begin{array}{lr}
\frac{\lambda}{\epsilon}(-m+\sqrt{m^{2}+\lambda^{2}})
+\arctan(\frac{m}{\epsilon})-\arctan(\frac{\sqrt{m^{2}
+\lambda^{2}}}{\epsilon}) & \mbox{if } m>0\\
\\
\frac{\lambda}{\epsilon}-\arctan(\frac{\lambda}{\epsilon}) & \mbox{if } m=0 \\
\\
\frac{\lambda}{\epsilon}(m+\sqrt{m^{2}+\lambda^{2}})
-\arctan(\frac{m}{\epsilon})-\arctan(\frac{\sqrt{m^{2}
+\lambda^{2}}}{\epsilon}) & \mbox{if } m<0.
\end{array}
\right.
\end{eqnarray}
We consider two cases: $m$ equal to zero (that is, the flux $\phi$ is zero) 
and $m$ different from zero. In the first case, after having removed the 
diamagnetic term, the conductivity is
\begin{equation}
\tilde{\sigma}_{xx}=\frac{e^{2}}{4h}\arctan(\frac{\lambda}{\epsilon}).
\end{equation}
Now we can take the limit for $\epsilon\rightarrow 0$ (thus making the 
cutoff irrelevant). We find that the conductivity for every node is different 
from zero and is equal to
\begin{equation}
\tilde{\sigma}_{xx}=\frac{1}{3}\frac{e^{2}}{4h}\frac{\pi}{2}.
\end{equation}
After the sum over all six nodes is done we can conclude that the longitudinal 
conductivity for $m=0$ is
\begin{equation}
\tilde{\sigma}_{xx}=\frac{e^{2}\pi}{4h}.
\end{equation}
The case where $m$ is not zero is quite different. Now, after having done the 
diamagnetic subtraction (\ref{subtr}), the conductivity is 
\begin{equation}
\tilde{\sigma}_{xx}=\frac{e^{2}}{4h}[\frac{m^{2}}{\epsilon}
\arccos(\frac{m}{\sqrt{m^{2}+\epsilon^{2}}})-\frac{m}{\epsilon}
+\arctan(\frac{m}{\epsilon})-\arctan(\frac{\sqrt{m^{2}
+\lambda^{2}}}{\epsilon})],
\end{equation}
but the limit for $\epsilon\rightarrow0$ gives a vanishing result.
We conclude that the longitudinal conductivity is different from zero 
only when there is no gap between the two bands, that is in our case for 
$m=0$. 

Next we consider the longitudinal conductivity for the honeycomb lattice. 
We expect that it is different from zero when the bands touch. This happens 
in six points, when $M=0$ and $\phi=0$, and in three points when 
$M=3\sqrt{3}\alpha t_{2}\sin(\phi)$. To calculate $\sigma_{xx}$, we use the 
expression (\ref{sigmaxx}). In the first case every node gives a contribution 
different from zero and equal to each other, so with the same observations 
made for the Kagome lattice we find that the longitudinal conductivity is
\begin{equation}
\sigma_{xx}=\frac{e^{2}\pi}{4h}\ \ \ (M=\phi=0).
\end{equation}
When $M>0$ and $\phi=0$ we find that the conductivity is
\begin{eqnarray}
\sigma_{xx}=&-&\frac{1}{2\hbar\epsilon}\frac{9}{8}
\frac{1}{(2\pi)^{2}}\int_{-\infty}^{\infty}\frac{\sqrt{h_{1}^{2}+h_{2}^{2}
+M^{2}}}{h_{1}^{2}+h_{2}^{2}+M^{2}+\epsilon^{2}}d^{2}k
\nonumber\\
&-&\frac{1}{2\hbar\epsilon}\frac{9}{8}\frac{1}{(2\pi)^{2}}
\int_{-\infty}^{\infty}\frac{M^{2}}{\sqrt{h_{1}^{2}+h_{2}^{2}+M^{2}}
(h_{1}^{2}+h_{2}^{2}+M^{2}+\epsilon^{2})}d^{2}k\nonumber
\end{eqnarray}
but, as in the case of the Kagome lattice, this integral vanishes; so in this 
case the longitudinal conductivity is zero.

The last case we consider is for $M=\pm 3\sqrt{3}\alpha t_{2}\sin(\phi)\ne0$. 
Here the bands touch only in three points and these give a contribution to 
the conductivity since now there is a gap where before there were three nodes. 
Around the three nodes the term $c$ is zero, so formally the problem is the 
same as that of the case $M=0$ and $\phi=0$ and only in these three points 
the longitudinal conductivity is different from zero. The result is equal to 
half of what was found in the case in which $M$ and $\phi$ are zero, that is
\begin{equation}
\sigma_{xx}=\frac{e^{2}\pi}{8h}\ \ \ 
(M=\pm 3\sqrt{3}\alpha t_{2}\sin(\phi)\ne0).
\end{equation}
We can conclude that the longitudinal conductivity can be rewritten as
\begin{equation}
\sigma_{xx}=\mu\frac{e^{2}\pi}{8h}
\end{equation}
with $\mu$ = 0, 1, 2 for the honeycomb lattice and $\mu$ = 0, 2 for the Kagome 
lattice.
  
\section{ Discussion and Conclusions }

We have considered both Haldane's model for electrons in a staggered flux 
on the honeycomb lattice and the model by Ohgushi, Murakami and Nagaosa for
electrons in the presence of a canted spin-1/2 texture on the Kagome lattice.
We have shown how similar these two models are in that the transverse Hall
conductivity $\sigma_{xy}$ is quantized as $\pm e^2/h$ as a function of the
tuning parameter (e.g. the magnetic flux per plaquette, $\phi$). 

Whilst for the Hall conductivity $\sigma_{xy}$ we have obtained the same 
results both for the Kagome lattice in the presence of a spin texture (as 
found by Oshgushi et al. (2000)) and for the honeycomb lattice with 
staggered magnetic field (as found by Haldane (1988)), we stress that 
we have used a different method of calculation based on implementing the band 
structure of each model in the Kubo formula. Furthermore, we have explicitely
evaluated in this way, and for the first time, also the longitudinal 
conductivity $\sigma_{xx}$ starting from the Kubo formula. This quantity is
also quantized in the absence of symmetry-breaking, non-ideal features of the 
system, but not in terms of integer multiples of $e^2/h$.    
 
For the Kagome lattice model, we find metallic behavior for a ferromagnetic 
state of localized spins perpendicular to the plane of the lattice. This 
state has a vanishing flux ($\phi=0$) in each plaquette of the Kagome lattice. 
Metallic behavior exists also for a canted state where the spins are inside 
the plane ($\theta=\pi/2$, in this case the local flux is $\phi=\pi$). The 
longitudinal conductivity is for both cases $\sigma_{xx}=e^2\pi/4h$ and the 
Hall conductivity vanishes. In Fig. 10 we show the schematic behaviour of the 
Hall conductivity for the model defined on the Kagome lattice and as a 
function of the parameter $\phi$. This is to be compared with the richer 
phase diagram for the Haldane model on the honeycomb lattice, reported in 
Fig. 11 also as a function of $\phi$. The longitudinal conductivity 
$\sigma_{xx}$ as evaluated in this work is shown schematically in Fig. 12.

\begin{figure}
\begin{center}
\includegraphics[width=9cm]{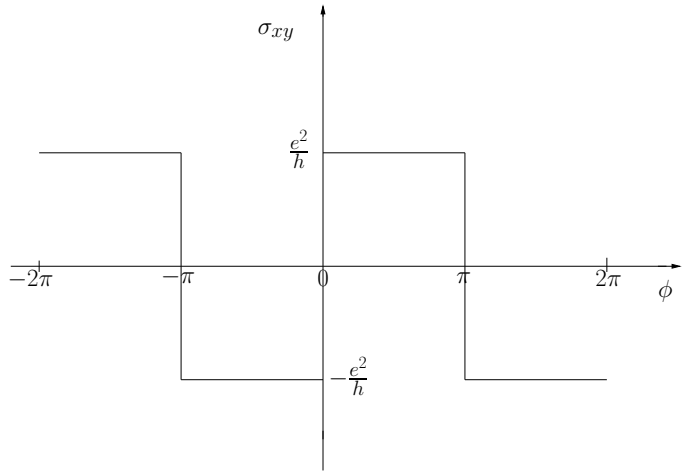}
\end{center}
\caption{Hall conductivity for electrons on the Kagome lattice.}
\end{figure}
                                                                                
\begin{figure}
\begin{center}
\includegraphics[width=9cm]{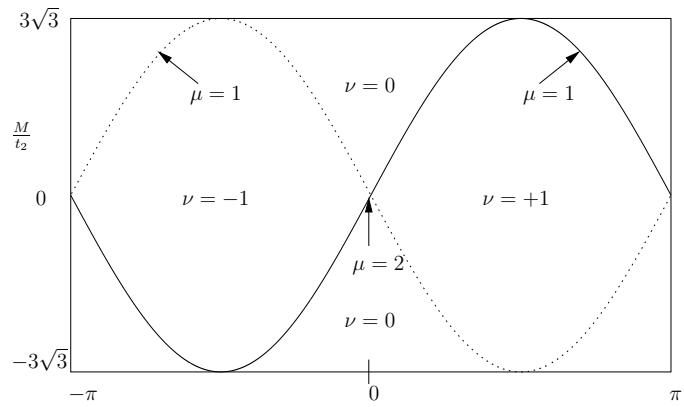}
\end{center}
\caption{Hall conductivity on the honeycomb lattice as a function of $\phi$;
the parameter $\nu$ quantizes $\sigma_{xy}$ in units of $e^2/h$ and 
$\mu$ quantizes $\sigma_{xx}$ in units of $e^2\pi/8h$.}
\end{figure}
                                                                                
\begin{figure}
\begin{center}
\includegraphics[width=12cm]{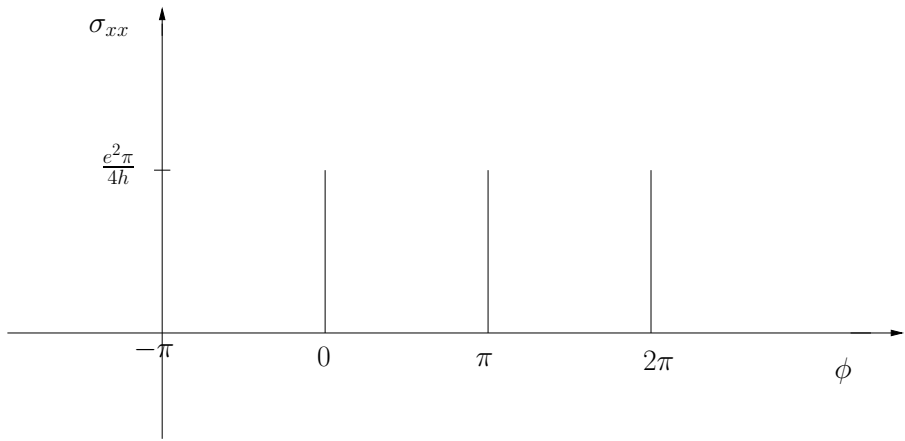}
\end{center}
\caption{Longitudinal conductivity for electrons on the Kagome lattice.}
\end{figure}

Removing some of the nodes in the DOS by breaking symmetries (like for the 
case of a square lattice with next-nearest neighbor terms) alters the Hall 
conductivity substantially. Also the introduction of disorder (e.g. slow 
fluctuations of the localized spins, fluctuations around the perfect canted
spin texture) may remove some of the nodes and yield non-universal features
in the transport properties. There is also another interesting effect due to 
disorder in our two-dimensional lattices. The longitudinal conductivity 
$\sigma_{xx}$ is usually based on diffusion of charge carriers; however, the 
diffusion coefficient $D$ is infinite in our model, since there is no 
scattering in the absence of imperfections. Nevertheless, the longitudinal 
conductivity, expressed through the Einstein relation
\[
\sigma_{xx}={e^2\over\hbar}D\rho ,
\]
is finite thanks to a vanishing density of states at the nodes. The
cancellation of the divergent diffusion coefficient and the vanishing
density of states is subtle. Since there is scattering by impurities in
a realistic system, a finite diffusion coefficient is more natural. On the
other hand, impurities create additional states near the nodes such that
a non-vanishing density of states exists. This effect was studied in the 
case of 2D Dirac fermions with random scatterers (Ziegler, 1997, 1998, Ziegler and Jug, 1997). In 
particular, it was found that random scattering broadens the metallic state 
(Ziegler and Jug, 1997), and the maximal conductivity value is lowered by a factor 
$1/(1+g/2\pi)$, where $g$ is the strength of the random fluctuations.

\vskip0.3cm
\noindent
Acknowledgement:

\noindent
We are grateful to MIUR (Ministero dell'Istruzione, Universita' e
della Ricerca) for support through PRIN-2003 grant and to
the Deutsche Forschungsgemeinschaft for support through
Sonderforschungsbereich 484. 

\section*{Appendix A: Linear Response and Kubo Formula}

From the Kubo formula we know that the conductivity tensor can be written as
(Madelung, 1978)
\begin{equation}
\sigma_{\mu\nu}=\frac{e}{i\hbar}\lim_{\alpha\rightarrow0}
\int_{-\infty}^{0}e^{(i\omega+\alpha)t}
Tr([\rho_{0},r_{\mu}]e^{-iHt/\hbar}j_{\nu}e^{iHt/\hbar})dt
\end{equation}
where $\rho_{0}$ is the Fermi function. Using the Green functions defined as
\begin{equation}
G_{\pm}(E)=(H/\hbar+E\pm i\eta)^{-1}
\end{equation}
we can use the substitution
\begin{equation}
e^{\pm iHt/\hbar}=\pm\lim_{\eta\rightarrow 0}
\int_{-\infty}^{\infty}e^{\mp iEt}G_{\mp}(E)\frac{dE}{2\pi i}\qquad (t\leq 0)
\end{equation}
the conductivity can be rewritten as
\begin{equation}
\sigma_{\mu\nu}=\frac{e}{\hbar}\int_{-\infty}^{\infty}
Tr{[\rho_{0},r_{\mu}]G_{+}(E)j_{\nu}G_{-}(E+\omega)}\frac{dE}{2\pi i}
\end{equation}
The current operator is
\begin{equation}
j_{\nu}=\frac{e}{i}[H,r_{\nu}]
\end{equation}
Using this expression iteratively, we find that the conductivity can be
rewritten as a sum of three terms
\begin{eqnarray}
\sigma_{\mu\nu}=-\frac{1}{\hbar}\int_{-\infty}^{\infty}
Tr\{\rho_{0}G_{+}(E)j_{\mu}G_{+}(E)j_{\nu}G_{-}(E+\omega)\}
\frac{dE}{2\pi} \nonumber\\
-\frac{1}{\hbar}\int_{-\infty}^{\infty}
Tr\{\rho_{0}G_{+}(E)j_{\nu}G_{-}(E+\omega)j_{\mu}G_{-}(E+\omega)\}
\frac{dE}{2\pi} \label{kubo}\\
+\frac{e}{\hbar}\int_{-\infty}^{\infty}
Tr\{\rho_{0}G_{+}(E)[j_{\nu},r_{\mu}]G_{-}(E+\omega)\}
\frac{dE}{2\pi}. \nonumber
\end{eqnarray}

\end{document}